\documentclass[onecolumn,floatfix,superscriptaddress,showpacs,showkeys,nofootinbib]{revtex4}%
\usepackage{epsfig}
\usepackage{amssymb,latexsym,amsmath}
\newcommand{\eq}[1]{\begin{align} #1 \end{align}}

\usepackage{bm}
\usepackage{amsmath}
\usepackage{empheq}
\usepackage{graphicx}
\usepackage{subfigure}
\usepackage{enumerate}
\usepackage{soul}
\usepackage[colorlinks=true,linktocpage=true,linkcolor=blue,citecolor=blue,allcolors=blue]{hyperref}
\usepackage[usenames,dvipsnames]{color}
\usepackage{url}

\begin{document}

\title{Estimation of the freeze-out parameters reachable in the AFTER@LHC project}
\author{Viktor Begun} \email{viktor.begun@gmail.com}
\affiliation{Warsaw University of Technology, Faculty of Physics, Koszykowa 75, 00-662 Warsaw Poland}

\author{Daniel Kiko\l{}a} \email{danie.kikola@pw.edu.pl}
\affiliation{Warsaw University of Technology, Faculty of Physics, Koszykowa 75, 00-662 Warsaw Poland}

\author{Volodymyr Vovchenko} \email{vlvovch@gmail.com}

\affiliation{Institut f\"{u}r Theoretische Physik, Goethe Universit\"{a}t Frankfurt, D-60438 Frankfurt am Main, Germany}
\affiliation{Frankfurt Institute for Advanced Studies, Giersch Science Center, D-60438 Frankfurt am Main, Germany}

\author{Daniel Wielanek} \email{daniel.wielanek@cern.ch}

\affiliation{Warsaw University of Technology, Faculty of Physics, Koszykowa 75, 00-662 Warsaw Poland}

\begin{abstract}
Estimation of the freeze-out parameters as the function of rapidity in Pb+Pb collisions at $\sqrt{s_{NN}}=72$~GeV in the AFTER@LHC project is performed.
The conventional hadron resonance gas model is used for the analysis of the events generated by the UrQMD model.
The results indicate that one may obtain at least $2.5$ times increase of baryon chemical potential $\mu_B$ in the forward rapidity range as compared to the mid-rapidity.
The $\mu_B$ values in the rapidity range of $0 < y < 4.5$ for AFTER@LHC are comparable to that covered by the RHIC Beam Energy Scan program.
Thus, a rapidity scan in the AFTER@LHC project provides a complementary approach to study the QCD phase diagram.
\end{abstract}


\keywords{heavy-ion collisions, QCD phase diagram, rapidity scan, LHC,
AFTER@LHC, LHCb, ALICE.}

\date{\today}

\maketitle

\section{Introduction}\label{Sec-Intro}
The determination of the phase structure of the strongly-interacting QCD matter is one of the most fundamental open questions in nuclear physics. The QCD phase diagram is usually expressed in terms of the temperature $T$, and the baryon chemical potential $\mu_B$. The experimental data collected in high-energy heavy-ion collision experiments at the Super Proton Sycnchrotron (SPS), at Relativistic Heavy-Ion Collider (RHIC) and at the Large Hadron Collider (LHC) provide a strong evidence that a state of matter with partonic degrees of freedom (the Quark-Gluon Plasma, QGP) is created in such collisions~\cite{Afanasiev:2002mx,Adcox:2004mh,Back:2004je,Arsene:2004fa,Adams:2005dq,Abelev:2014pua,ATLAS:2011ah, Aamodt:2010jd,Aad:2010bu, Chatrchyan:2011sx}. The lattice QCD calculations indicate that there is a ''smooth'' phase transition between the hadronic matter and the QGP at $\mu_B = 0$,  the so-called crossover transition~\cite{Aoki:2006we,Borsanyi:2010cj,Bazavov:2011nk}.
Theory calculations suggest that in the range of large $\mu_B$ the 1st order phase transition takes place~\cite{Klevansky:1992qe,Barducci:1993bh,Stephanov:1996ki,Alford:1997zt,Stephanov:1998dy}, thus the QCD critical point, that separates these two types of the phase transitions, is expected to exist. However, the corresponding theoretical calculations provide little guidance on the position of the critical point on the temperature and baryon chemical potential phase diagram - the predictions cover almost the whole $T(\mu_B)$ plane~\cite{Stephanov:2004wx}.
Therefore, the search for the critical point of the QCD matter is one of the main motivations for the ongoing and future experiments on heavy ion collisions (see, e.g., Refs.~\cite{Keane:2017kdq,STAR:BES2,Turko:2018kvt,Golovatyuk:2016zps,Ablyazimov:2017guv, Sako:2015cqa}).

The present experimental methods to change $T$ and $\mu_B$ are either to vary the collision energy, or to change the types of colliding nuclei.
However, these approaches have some significant constraints.
The available energies are limited by the accelerator capabilities, see for example the  RHIC Beam Energy Scan program phase I and phase II~\cite{STAR:BES2}. The existing observables might be sensitive to the critical point only in it's closest vicinity, see e.g. Ref.~\cite{Antoniou:2017vti}. Thus, it is possible to miss the critical point in a collision energy scan.
Moreover, the amount of the available statistics decreases significantly with decreasing bombarding energy in the collider mode.
This limits the high-statistics studies to low transverse momenta and only to the most abundant particle species.

In this paper we recall the alternative approach: varying the rapidity interval at a fixed collision energy~\cite{Dumitru:1993vz,Biedron:2006vf,Broniowski:2007mu,Becattini:2007ci,Becattini:2007qr}. 
The topic recently received a renewed interest, the physics discussed include high baryon densities achievable at the LHC and RHIC in the fragmentation region~\cite{Li:2016wzh,Ivanov:2018rrb},
or searching for critical point signatures through the rapidity dependence of proton number cumulants \cite{Brewer:2018abr}.
In a rapidity scan one compares the subsystems corresponding to different rapidity bins and characterizes them by different thermodynamic parameters.
In the simplest corresponding physical picture the rapidity axis is populated at freeze-out by fireballs, each characterized by thermal parameters, which depend on fireball's space-time rapidity $y_{_{FB}}$~\cite{Becattini:2007qr}. All fireballs contribute to the hadron yield spectrum at a given rapidity $y$. However, the dominant contributions do come from fireballs with $y_{_{FB}} \simeq y$.
Therefore, the $y_{_{FB}}$ dependence of thermal parameters at freeze-out translates into the $y$ dependence of the final state hadron chemistry.
We provide estimates of the $T$ and $\mu_B$ values reachable at different rapidities in a fixed target experiment at the LHC - the AFTER@LHC project~\cite{AFTERatLHC}~(see also~\cite{KikolaQM}), by analyzing the hadron yield chemistry in different rapidity bins using the UrQMD model\footnote{The UrQMD model has been used before to estimate temperature and baryo-chemical potentials reached during fireball evolution in different space-time domains~\cite{Bravina:1998pi,Bravina:1998it,Bravina:1999dh,Endres:2014zua}.
These analyses were done for the whole time evolution rather than for the freeze-out as in this work.}~\cite{Bass:1998ca,Bleicher:1999xi}.
We compare the outcome with the results of the analysis of the total rapidity range, and with the energy scan at fixed rapidity $y\sim 0$.
The goal of this paper is to answer the question whether a rapidity scan at the AFTER@LHC can be used to change $T$ and $\mu_B$ by the amount notably larger than the uncertainty of $T$ and $\mu_B$, and comparable to the amount reachable in an energy scan at mid-rapidity.

For the measurement of particle yields at larger rapidities proposed in this paper, the ALICE detector has to successfully operate in a fixed-target mode. Even though ALICE was designed as a collider experiment, it is well suited for measurements of fixed-target collisions. The ALICE main detector, the Time Projection Chamber (TPC), has an excellent spatial resolution, and it is capable of measuring thousands of particle tracks in a collision. Therefore, it is an appropriate tool for recording fixed-target events, where most of the particles fly together due to large boost. 

The STAR experiment provides an example that this is indeed the case. The STAR at the Relativist Heavy Ion Collider is similar to the ALICE detector in many aspects: it has a similar geometry, the TPC is the main tracking device, and there is the Time-of-Flight detector installed outside the TPC to improve particle identification. STAR has performed successful fixed-target measurements of Au-Au collisions using these two detectors for particle track and primary vertex reconstruction, and the particle identification~\cite{Meehan:2016qon,Meehan:2017cum}. STAR uses a thin gold foil placed at the entrance to the STAR TPC to study collisions at the low-energy range (3 $< \sqrt{s_{NN}}< $ 7 GeV), where the operation in a collider mode was unfeasible or unpractical due to a low collision rate. The STAR Fixed Target program extends the $\mu_B$ range available in RHIC BES and provides good quality data that can serve as reference measurements for the search of the phase transition and the critical point. Following its original success, the fixed-target mode is a part of the Beam Energy Scan II program, planned for 2019 and 2020. The positive experience of STAR provides a proof that detector designed for central-rapidity measurement in the collider mode can work very well in a fixed-target mode.

A fixed target experiment using TeV beams of the Large Hadron Collider exhibits a couple of unique features. Such an experiment will have a wide kinematic coverage. The collision of 2.76 TeV heavy-ion (for instance Pb) beam on a fixed target releases a center-of-mass-system (c.m.s.) energy per nucleon pair $\sqrt{s_{NN}} = 72$~GeV.
These collisions occur in a frame with an extremely large Lorentz factor $\gamma \approx 4.3$. Due to the Lorentz boost, the forward instrumentation of existing LHC detectors provides a mid-rapidity coverage in the c.m.s.
For example, the LHCb experiment with the pseudorapidity coverage of $2 < \eta < 5$ in the laboratory frame would cover the c.m.s. rapidity of $-2.3 \lesssim y \lesssim 0.7$. In the case of the ALICE experiment, the central detector (with $|y^{Lab}|<0.9$) becomes a ''backward'' one in the fixed-target mode, with access to the large absolute value of rapidity in the c.m.s. with $-5.2 \lesssim y^{c.m.s.} \lesssim -3.4$. In general, AFTER@LHC will provide a broad kinematic coverage $-5.2 \lesssim y^{c.m.s.} \lesssim 0.7$, which is not the case in the collider mode.

Particle identification at large rapidity is challenging. However, prior studies have proven that the LHCb detector has exceptional performance in identifying various hadrons in a broad rapidity range.
More specifically, the LHCb collaboration has reported measurements of identified yields in p+p collisions at $\sqrt{s}= 0.9$ and 7 TeV in rapidity range of $2 < y \lesssim 4.5$ for proton/antiproton, $K^+/K^-$ and $\pi^+/ \pi^-$ yield ratios, see Tables 8-10 in~\cite{Aaij:2012ut}, as well as the $K^0_S$ and $\Lambda$ production measurements in~\cite{Aaij:2011va}.
More recently, the LHCb collaboration reported also the preliminary results of the antiproton production measurement in a p+He collision in a fixed target mode, at momenta corresponding to large rapidities~($y \simeq 4-5$), see~\cite{Maurice:2017iom}, and~\cite{LHCb:2017tqz}. These experimental results prove that the LHCb is capable of effective hadron identification over a rapidity range of at least $2 \lesssim y \lesssim 5$. 

The available luminosity per year is very high, similar to the nominal LHC luminosities, and orders of magnitude larger than that at RHIC in the similar energy range~\cite{Massacrier:2017oeq}.
Such amount of data gives access to rare probes like heavy quarks -- quarkonium~\cite{Massacrier:2015qba} and open heavy flavor hadrons.
In addition, one can easily change the atomic mass of the target, which facilitate the system-size dependence study.
In general the AFTER@LHC project will have a broad physics program which includes the high-$x$ frontier, the spin physics at the LHC (given the installation of the polarized target) and the heavy-ion physics (see, e.g.,~\cite{Brodsky:2012vg, Massacrier:2017oeq, Kikola:2017hnp, Trzeciak:2017csa, Kikola:2015lka} and other publications and presentations in \cite{AFTERatLHC}).
These goals can be achieved in a cost effective way, by collecting data in the fixed-target mode in the existing experiments (LHCb and ALICE) in parallel to the collider-mode operation.
For the study described in this paper, we assume that AFTER@LHC will deliver a sufficiently good particle identification over a broad rapidity range. This expectation is based on the capabilities of the LHCb and the ALICE detectors.
Moreover, the LHCb experiment has already proven that it is capable of registration and identification of particles, and performing physics measurements in a fixed-target mode~\cite{Maurice:2017iom}.

\section{Rapidity scan vs energy scan}\label{Sec-I}
\begin{figure}[h!]
\includegraphics[width=0.49\textwidth]{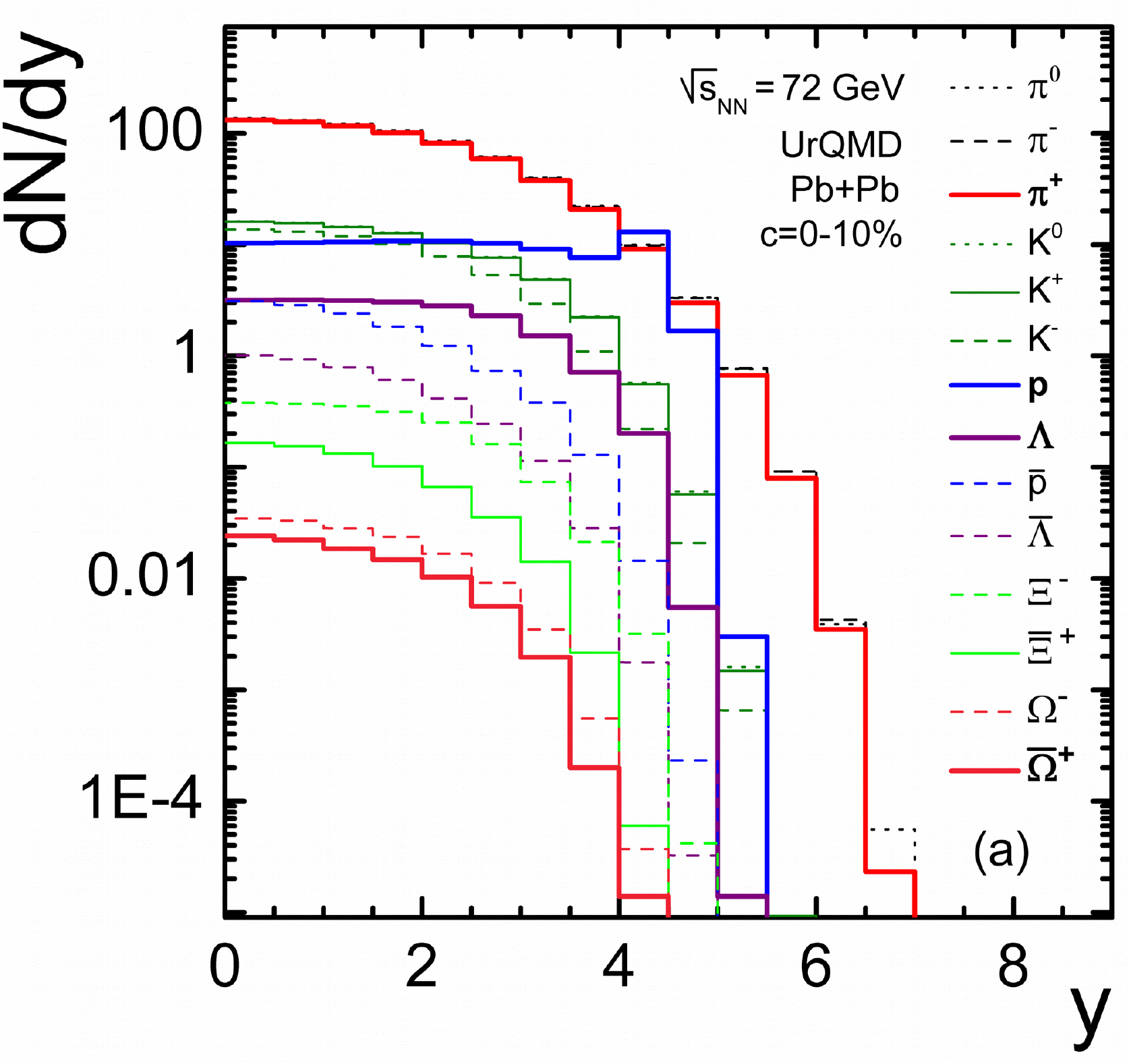}~~~~
\includegraphics[width=0.49\textwidth]{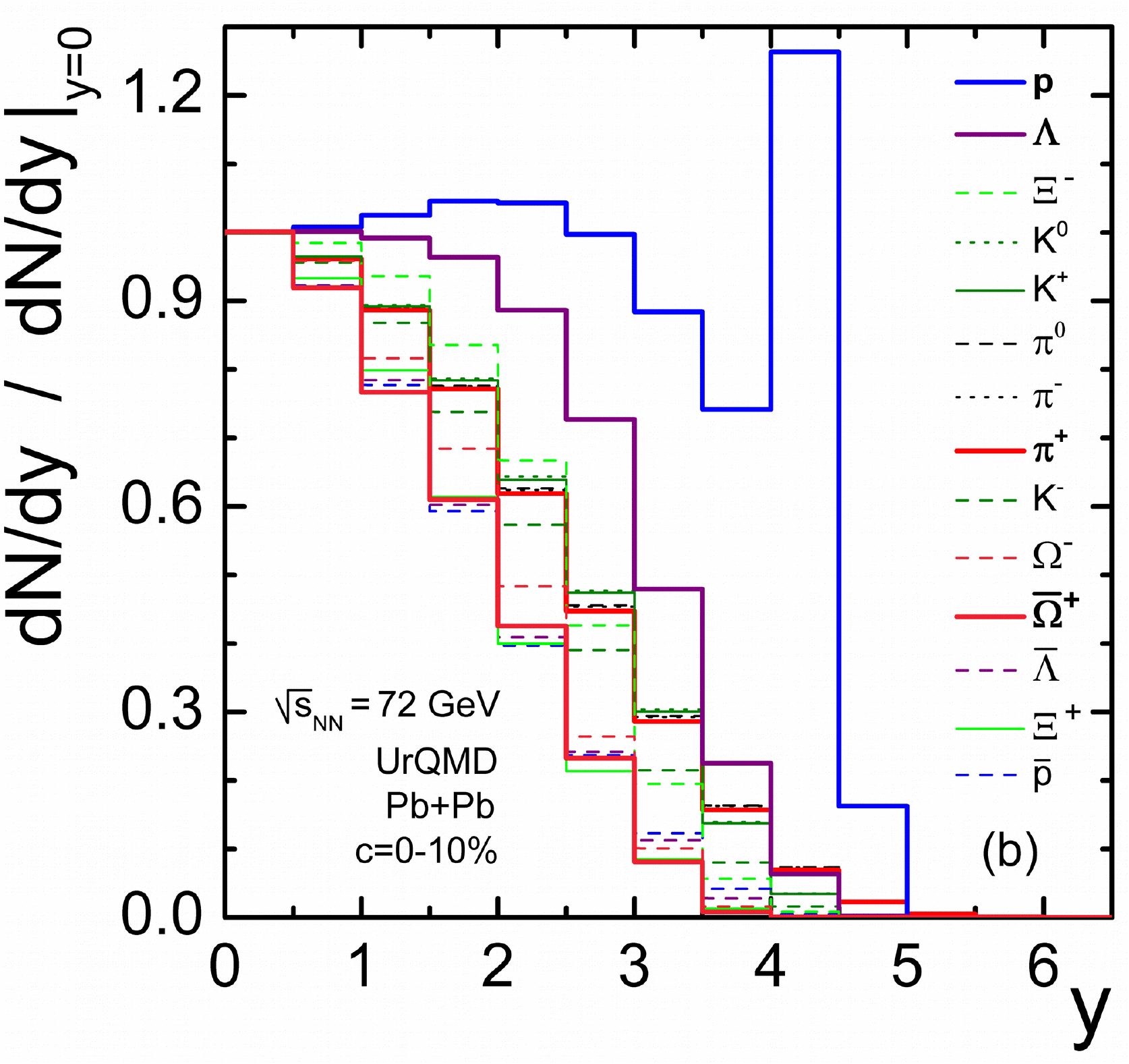}
\caption{(a) Rapidity dependence of particle multiplicities per event generated in the UrQMD model.
The order of particle labels is sorted according to their abundance at $y=0$. (b) The same as (a), but normalized to the corresponding value at mid-rapidity. The particle labels in (b) are sorted according to their abundance in the rapidity interval $2<y<2.5$.}
\label{fig-dNdy}
\end{figure}

We perform the estimates of the temperature $T$, baryo-chemical potential $\mu_B$, and volume $V$ of the analyzed systems in two steps.
First, we calculate the mean multiplicities of produced particles using the UrQMD model~\cite{Bass:1998ca,Bleicher:1999xi}.
Second, we fit the obtained mean multiplicities in a hadron resonance gas (HRG) model.
We use the UrQMD-3.4 model to generate the $10\%$ most central Pb+Pb collisions at the $\sqrt{s_{NN}}=72$~GeV. The number of events is
$N_{ev}\simeq 2*10^5$.
The UrQMD is run in the cascade mode, without hydro, for simplicity.
The resulting rapidity distribution for different hadron species with the rapidity steps $\Delta y=0.5$ is shown in Fig.~\ref{fig-dNdy}.
The $\Delta y$ step is chosen so that it is small enough to catch the forward rapidity increase of the $dN/dy$ for protons, but not too small so as to have enough statistics for the chosen particles. The increase in $dN/dy$ for protons is very well seen in the data, see e.g. Refs.~\cite{Appelshauser:1998yb,Arsene:2009aa,Videbaek:2009zy,Anticic:2009wd}, and is reproduced by the UrQMD, see e.g. Refs.~\cite{Mitrovski:2008hb,Ozvenchuk:2017epq}.
The particle set is chosen in order to be able to constrain the HRG parameters in a thermal fit.

In our fits we use
the \texttt{Thermal-FIST}~\cite{thermal-fist} thermal model package,
previously used in Refs.~\cite{Vovchenko:2015idt,Begun:2018qkw}. A general description of a HRG model can be found following Refs.~\cite{Wheaton:2004qb,Torrieri:2004zz,Petran:2013dva}.
For the recent applications of a HRG see e.g.~\cite{Andronic:2017pug,Becattini:2017pxe,Chatterjee:2016cog}, and references therein.
We consider only the grand canonical ensemble, thus, we imply that baryon number $B$, electric charge $Q$, and strangeness of the system $S$, are conserved on average.
We assume no additional interactions between particles in a HRG, therefore excluded-volume~\cite{Rischke:1991ke,Yen:1997rv,Yen:1998pa}, or van der Waals interaction effects~\cite{Vovchenko:2016rkn} are omitted.
This provides an internal consistency of the procedure since these interactions are presently omitted in UrQMD as well.
We expect strangeness non-equilibrium in small $dN/dy$ bins, therefore, we do include the $\gamma_S$ parameter~\cite{Letessier:1993hi} into fits.
Thus, we have six parameters in the HRG - the system temperature $T$, the three chemical potentials $\mu_B$, $\mu_Q$, $\mu_S$, the $\gamma_s$, and the radius of the system $R$.
The radius $R$ is related to the system volume as $V=(4/3)\pi R^3$.
Two chemical potentials, $\mu_Q$ and $\mu_S$, are constrained for each considered rapidity bin by the conditions  of zero net strangeness $\langle S\rangle=0$, and by the ratio of electric charge to baryon charge of $\langle Q\rangle/\langle B\rangle\simeq 0.4$, as in Pb nuclei. Thus, the number of free parameters is four -- $T$, $\mu_B$, $\gamma_s$, and $R$.
The set of the input mean multiplicities should contain all three conserved charges, mesons, baryons, and their anti-particles.
Therefore we choose the following particle set, which satisfies these conditions: $\pi^0$, $\pi^{\pm}$, $K^0$, $K^{\pm}$, $p$, $\bar{p}$, $\Lambda$, $\bar{\Lambda}$, $\Xi^-$, $\bar{\Xi}^+$, $\Omega^-$, $\bar{\Omega}^+$.
A thermal fit requires both mean hadron yields and their uncertainties.
The statistical uncertainty scales with the number of events as $1/\sqrt{N_{ev}}$, and becomes very small for a large number of generated events.
Therefore, we additionally assume the theoretical {\it systematic} uncertainty at the level of 10\% for each UrQMD generated yield.
This assumption for uncertainties is comparable to the total experimental uncertainties reported in the RHIC BES~\cite{Adamczyk:2017iwn}, and in the ALICE experiment~\cite{ABELEV:2013zaa,Abelev:2013vea}. 


%
\begin{figure}[t]
\includegraphics[width=0.49\textwidth]{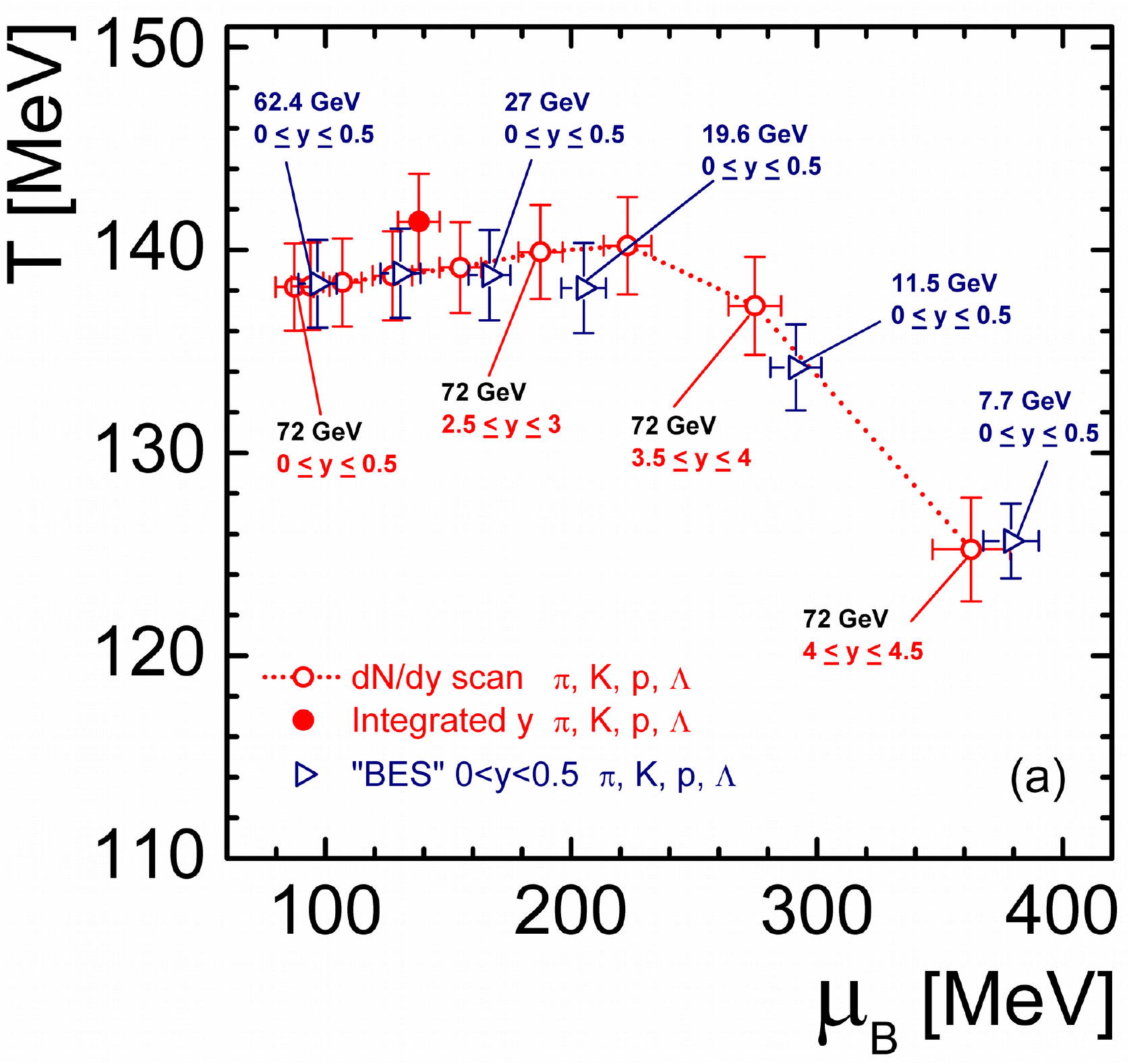}~~~~
\includegraphics[width=0.49\textwidth]{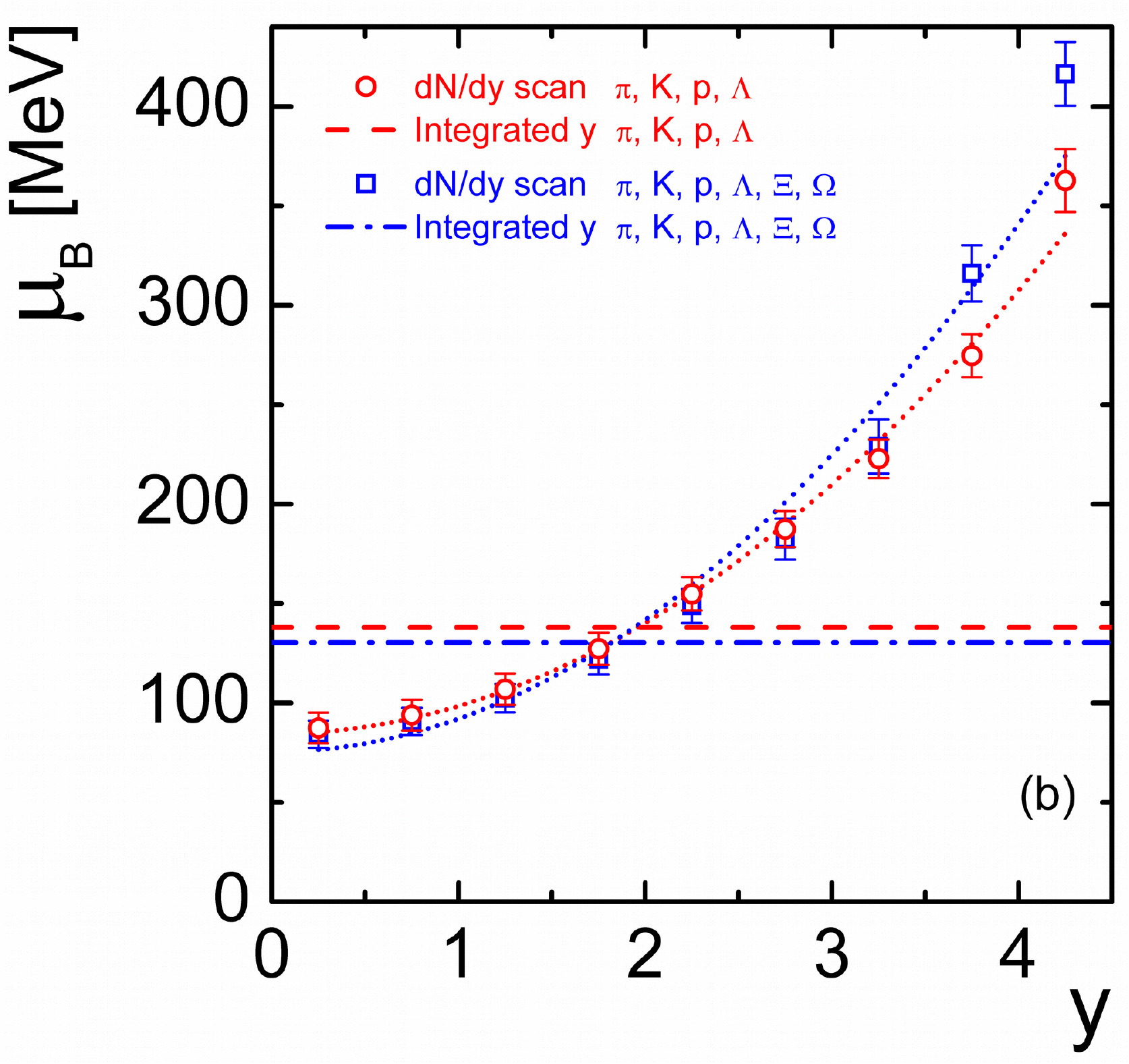}
\caption{(a) Rapidity scan at fixed energy vs energy scan at fixed rapidity
(b) The $\mu_B$ as the function of rapidity.
}
\label{fig-T-muB}
\end{figure}

Figure~\ref{fig-T-muB} (a) shows the $T$ and $\mu_B$ values obtained in the rapidity scan with the step $\Delta y=0.5$ at fixed energy ($\sqrt{s_{NN}}=72$~GeV, open circles), in comparison with the energy scan at the fixed rapidity ($0<y<0.5$, open triangles). The energy scan corresponds to the RHIC BES program~\cite{Adamczyk:2017iwn}. Note that we use the same mid-forward-rapidity interval $0<y<0.5$ in order to compare the energy scan and the rapidity scan in the UrQMD, while the RHIC BES results correspond to symmetric and narrower interval $-0.1<y<0.1$. In order to point out this difference we call our UrQMD beam energy scan ''BES''.
The $dN/dy$ yields used in the rapidity scan are those shown in Fig.~\ref{fig-dNdy}.
The particle set for the fit of ''BES'' was limited to $\pi^0,~\pi^{\pm},~K^0,~K^{\pm},~p,~\bar{p},~\Lambda,~\bar{\Lambda}$, i.e. excluding the heavy $\Xi^{-}$, $\bar{\Xi}^{+}$ and $\Omega^{-}$, $\bar{\Omega}^{+}$, see discussion of Fig.~\ref{fig-T-Chi2-y} below.
The $T$ and $\mu_B$ values obtained in the fit to the corresponding rapidity-integrated~($4\pi$) yields at $\sqrt{s_{NN}}=72$~GeV are shown by the full circle in Fig.~\ref{fig-T-muB} (a).
One can see that the integrated yields are similar to an average $T$ and $\mu_B$ obtained in the rapidity scan in the moderate rapidity interval $0<y<3$.
This is expected since the bulk of hadron production is peaked at $y=0$.
The rapidity scan over all rapidities covers almost the same $T$ and $\mu_B$ range as the energy scan at mid-rapidity.

Figure~\ref{fig-T-muB} (b) depicts $\mu_B$ as the function of rapidity. Triangles correspond to the whole particle set shown in Fig.~\ref{fig-dNdy}, while circles - to the set without $\Xi$ and $\Omega$ as in Fig.~\ref{fig-T-muB}~(a).
The results for the $\mu_B$ appear to be almost independent on the particle set. It happens, because the UrQMD was constrained to the rapidity distributions at various energies~\cite{Mitrovski:2008hb}.
The increase of $\mu_B$ with $y$ is caused by the data driven increase of the baryon number $dN/dy$ and by the decrease of the $dN/dy$ for other particles at large $y$, see Fig.~\ref{fig-dNdy} and Refs.~\cite{Appelshauser:1998yb,Arsene:2009aa,Videbaek:2009zy,Anticic:2009wd,Anticic:2016ckv}.
A similar increase of $\mu_B(y)$ was observed earlier at top RHIC energy of $\sqrt{s_{NN}}=200$~GeV, and was fitted with a parabola~\cite{Becattini:2007ci,Becattini:2007qr}:
 \eq{
 \mu_B(y)~=~a+b\,y^2~,
 }
where the fit parameters were found to be $a=25\div26$~MeV and $b=11\div12$~MeV.
We perform a similar parabolic fit to the extracted $y$-dependence of $\mu_B$  at the AFTER@LHC energy, see dotted lines in Fig.~\ref{fig-T-muB} (b).
We obtain $a=75.5\pm5.9$, $b=16.6\pm 1.0$ for the full particle set, and $a=84.5\pm3.2$, $b=13.9\pm0.5$ for the set without $\Xi$ and $\Omega$. The $a$ parameter is larger than in~\cite{Becattini:2007ci,Becattini:2007qr}, because we study a smaller collision energy, see~\cite{Cleymans:1998fq, Andronic:2005yp, Vovchenko:2015idt,Andronic:2017pug}.
The $b$ parameter is similar to~\cite{Becattini:2007ci,Becattini:2007qr}, with an indication for the stronger increase of $\mu_B$ at large $y$.
%

The chemical freeze-out temperature, $T(y)$, is approximately constant for the largest part of the rapidity interval considered, $y \lesssim 3$. 
Notable temperature changes appear only at larger rapidities, $y \gtrsim 3$, as seen in Fig. 3(a).
%
If the $\Xi$ and $\Omega$ UrQMD yields are considered in the thermal fit, then a pronounced peak in the $T(y)$ dependence is observed at $y \simeq 3.5$.
This is required to describe these UrQMD yields with the HRG model.
Analysis of the rapidity bin dependence shows that the  peak in $T(y)$ is seen if the rapidity step is small enough: $\Delta y\lesssim 2$.
If $\Delta y$ is decreased further below the $\Delta y = 0.5$ currently employed, than the peak for $T(y)$ in Fig.~3~(a) becomes slightly sharper.
When the relative amount of $\Xi$, $\Omega$, and other particles produced by UrQMD as the function of rapidity will be confirmed or rejected experimentally, then it would ultimately be possible to say whether the maximum in $T(y)$ is an artifact of the modeling, or a real effect.
Note that the notable temperature changes shown in Fig.~3(a) correlate with a fast increase of $\chi^2 / dof$ at $y \gtrsim 3$~[Fig. 3(b)], indicating that the interpretation of the corresponding large rapidity bins in terms of temperature and baryochemical potential may be questionable.
We note that in prior works it was either assumed that $T(y)\simeq const$~\cite{Biedron:2006vf,Broniowski:2007mu}, while the drop of $T$ has also been reported in~\cite{Becattini:2007ci,Becattini:2007qr}.
In contrast to our work, the growth of $T(y)$ with rapidity is not observed in~\cite{Becattini:2007ci,Becattini:2007qr}. 
The reason is that the authors of~\cite{Becattini:2007ci,Becattini:2007qr} have enforced a parabolic increase of $\mu_B(y)$, and also enforced that T is a unique function of $\mu_B$, such that the $T$-$\mu_B$ values always lie on the chemical freeze-out curve.  
It means that in~\cite{Becattini:2007ci,Becattini:2007qr} $T$ always monotonically decreases by construction, as long as $\mu_B$ increases. We do not employ such a constraint in the present work.

\begin{figure}[t]
\includegraphics[width=0.49\textwidth]{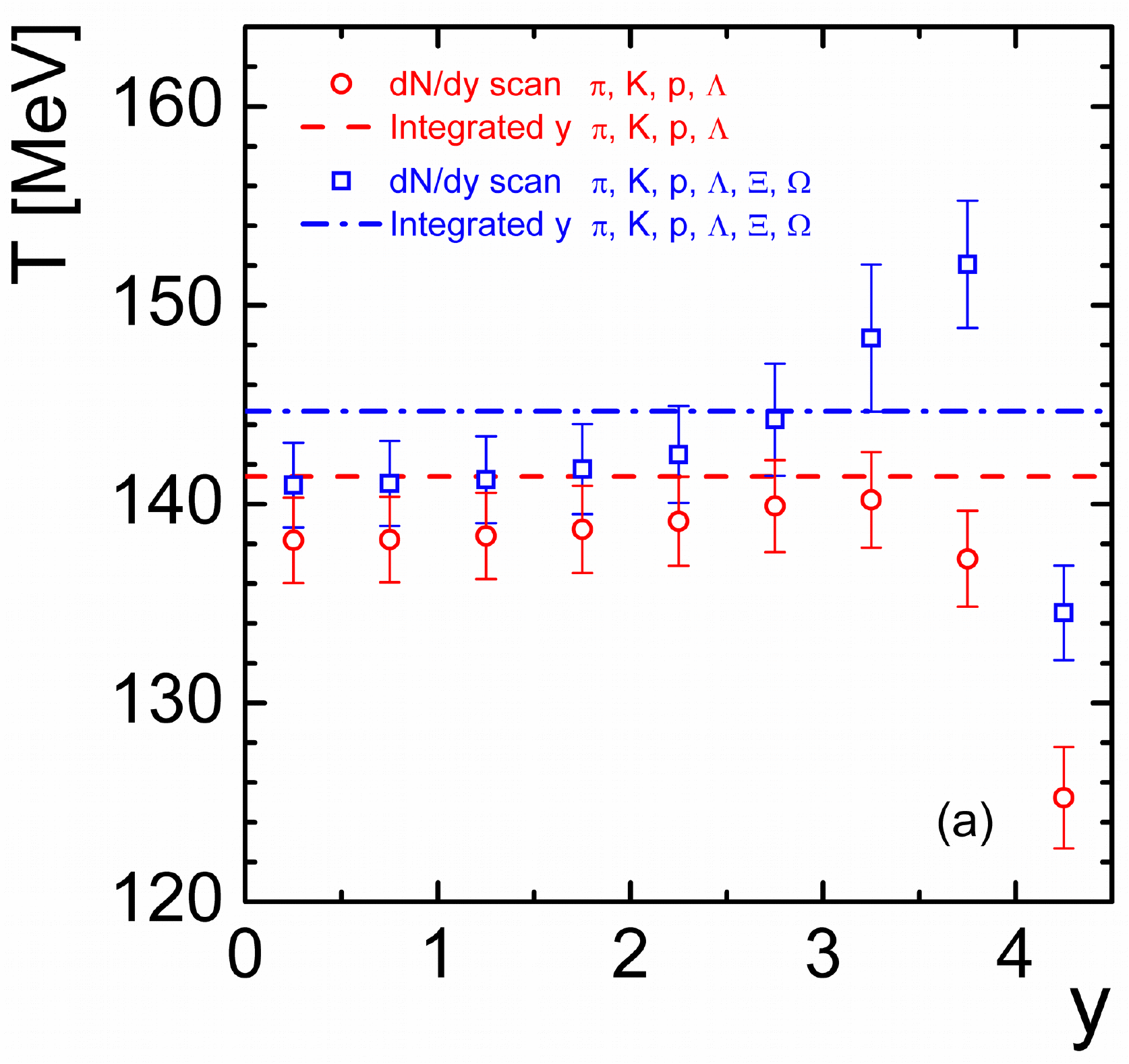}~~~~
\includegraphics[width=0.49\textwidth]{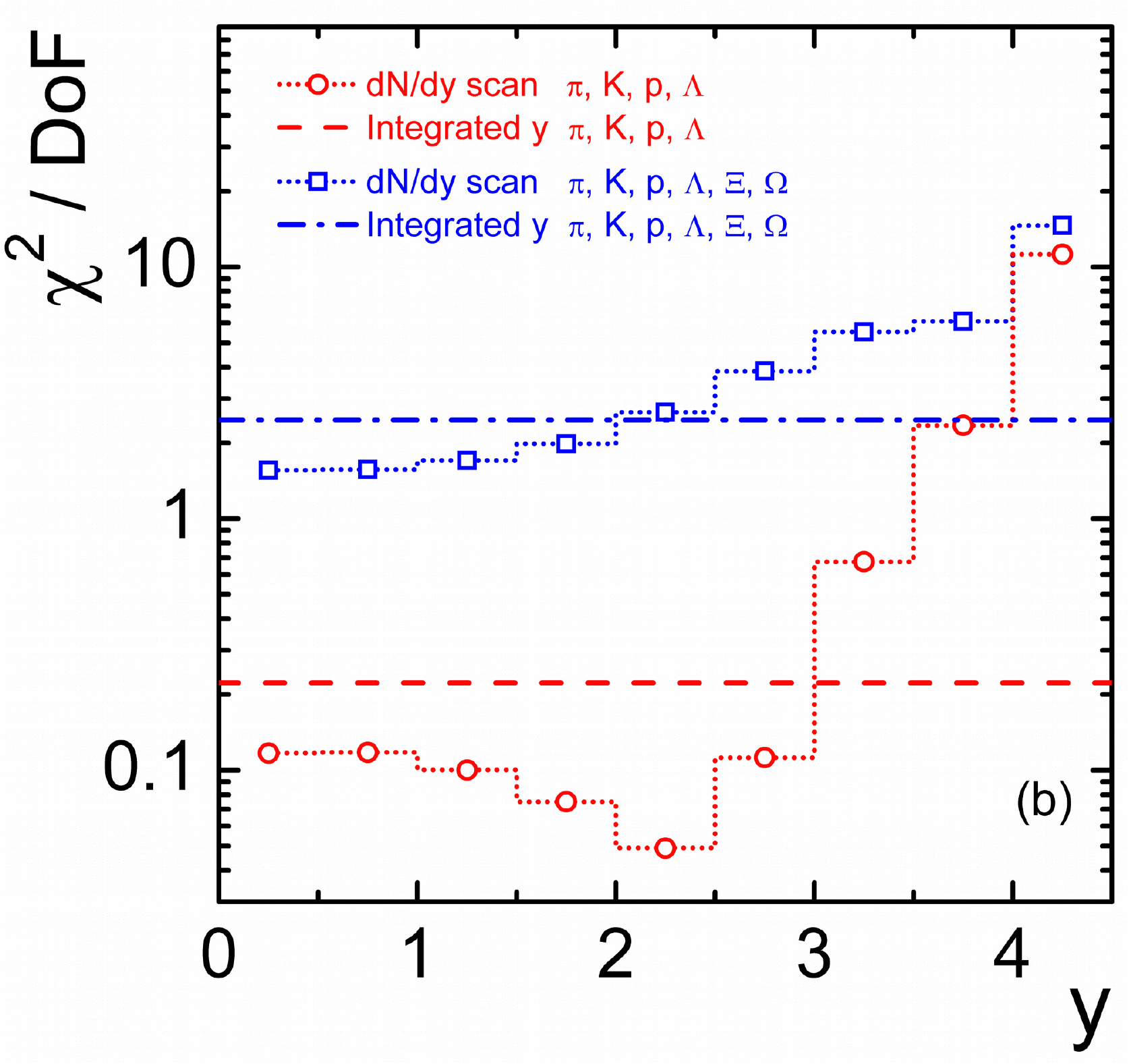}
\caption{The extracted chemical freeze-out temperature (a) and the $\chi^2$ per degree of freedom (b) as the function of rapidity.
}
\label{fig-T-Chi2-y}
\end{figure}

The obtained temperature values are about 20~MeV smaller than the temperatures extracted from fits to the real data at RHIC and LHC.
In that regard one should note that UrQMD contains string excitation and fragmentation, as well as many-body decays, which result in a Hagedorn-like equation of state in box simulations~\cite{Belkacem:1998gy,Zabrodin:2016ugs}.
The temperature values that we observe, $T \sim 140$~MeV, appear to be caused by this property of the UrQMD model. Correcting for this UrQMD behavior, one arrives at $T \sim 150-160$~MeV as an estimate of the chemical freeze-out temperatures, as expected for the AFTER@LHC project from the common $T(\mu_B)$ freeze-out line~\cite{Cleymans:1998fq, Andronic:2005yp, Vovchenko:2015idt,Andronic:2017pug}.

The uncertainties of the HRG parameters result from the assumed 10\% {\it systematic} uncertainties of the input multiplicities from the UrQMD.
The freedom in choosing the input uncertainties means that the absolute values of $\chi^2$ per degree of freedom depend on this assumption.
However, the rapidity dependence of $\chi^2/\rm DoF$ tell us how good is the treatment of the neighboring $\Delta y$ bins as single fireballs, see Fig.~\ref{fig-T-Chi2-y} (b), and the discussion in Appendix.
We do note that $\chi^2$ values are smaller at small rapidities, 
and show a rapid increase for large ($y>2.5$) rapidities.

The largest relative increase of $\mu_B(y)$ obtained here for the AFTER@LHC project is a factor of about 5.
This increase is reached between the smallest $0\lesssim y\lesssim 0.5$ and the largest $4.0\lesssim y\lesssim 4.5$ considered rapidity bins, and for the particle set with $\Xi$ and $\Omega$, where this effect is the strongest, see the square symbols in Fig.~\ref{fig-T-muB}~(b). 
However, the $\chi^2(y)$ of the thermal fit grows fast at $y\gtrsim 3$~[Fig.~\ref{fig-T-Chi2-y}~(b)], 
thus, the interpretation of these large rapidity bins in terms of the HRG model may be doubtful.
Therefore, let us denote the rapidity range, $0<y<3$, as the ''conservative'' one, and the full considered rapidity range, $0<y<4.5$, as the ''optimistic'' one.
The temperature in the ''conservative'' rapidity range is the same for both particle sets, while the $\mu_B$ changes by {\it at least} the factor of 2.5.
The change of the $\mu_B$ in the ''optimistic'' range is much higher.
%


We checked that the change of $\mu_B$ with rapidity can be expected to be even stronger in peripheral collisions, but also with a stronger dependence on the particle set used. The qualitative conclusion from the study of peripheral collision is similar. Therefore, we limit our study only to the case of the most central collisions for clarity.

The rapidity dependence of $\mu_B$ obtained here is determined by the $dN/dy$ spectra produced by UrQMD.  
We use the UrQMD model in the cascade  mode, without the intermediate hydro stage. 
Therefore, the main physics which determines the $dN/dy$ spectra is embedded in the string excitation and fragmentation mechanism implementation in UrQMD, and in the further transport of the baryon number and strangeness. 
There are many parameters which determine these mechanisms in UrQMD.
They were fixed by the UrQMD collaboration by fitting the various existing data at larger and at smaller energies in elementary (hadron-hadron) and heavy-ion collisions. 
We do not vary these parameters and use the default ones.
It is possible that the precise value of the relative increase of the baryon chemical potential can differ, if the UrQMD parameters are varied.
The variation of these parameters would require additional assumptions and could break existing agreement of UrQMD with data, therefore it is not considered here.

\begin{figure}[h!]
\includegraphics[width=0.49\textwidth]{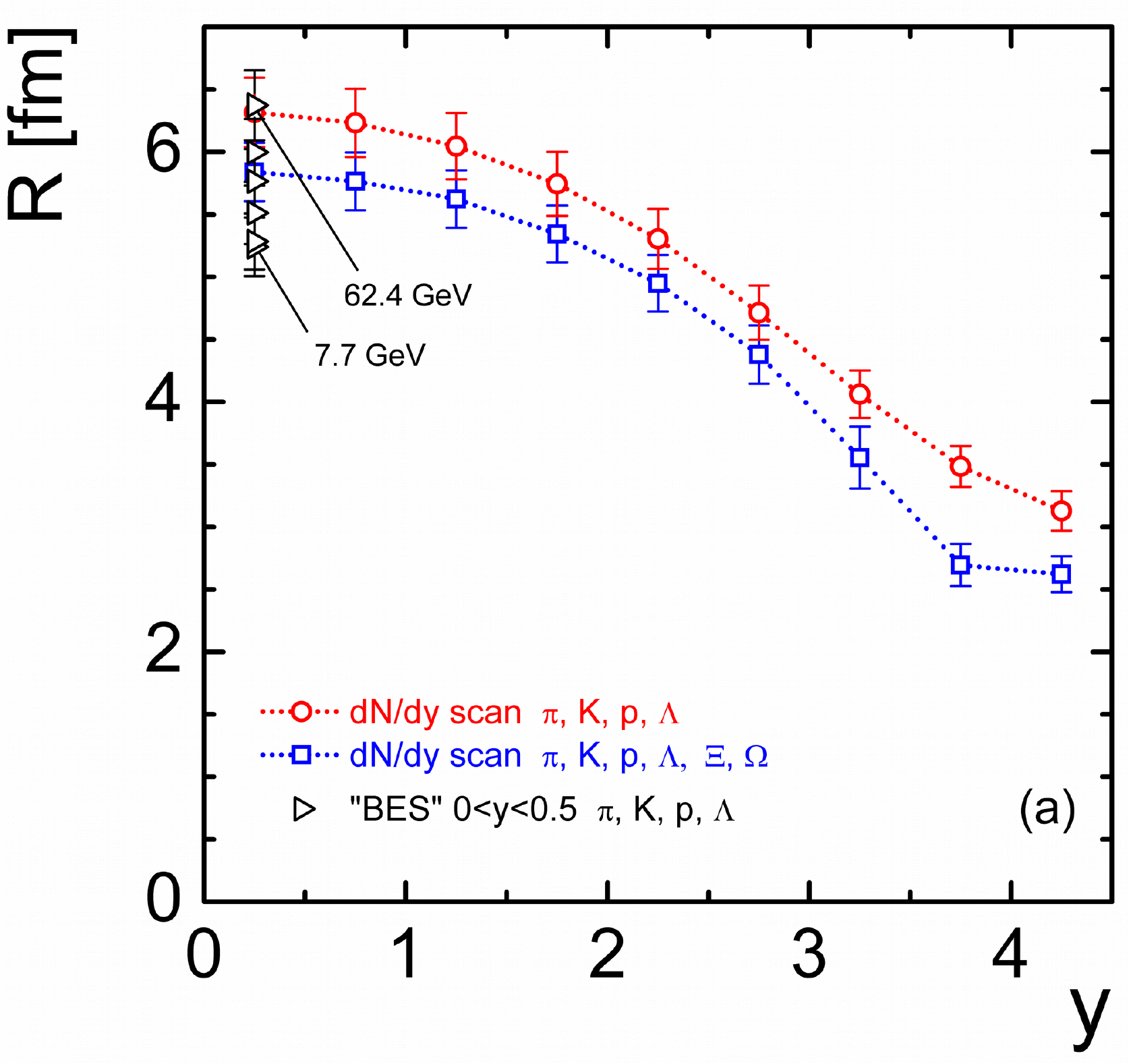}~~~~
\includegraphics[width=0.49\textwidth]{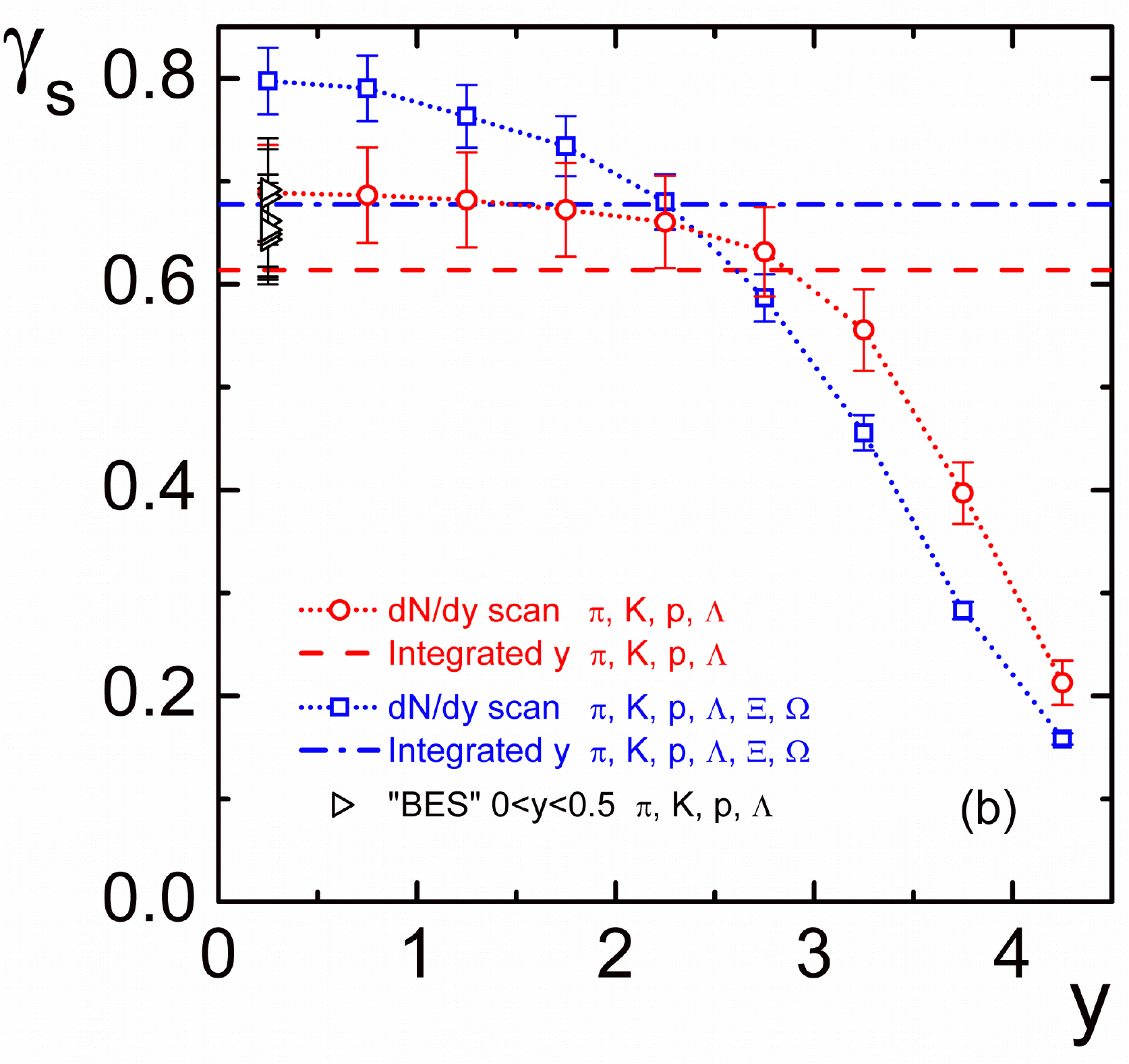}
\caption{(a) The system radius parameter $R$, and the strangeness saturation parameter $\gamma_S$ (b) as the function of rapidity.
}
\label{fig-R-gS}
\end{figure}

The radius $R$, and the strangeness saturation parameter, $\gamma_S$, behave as expected: they decrease with $y$, see~Fig.~\ref{fig-R-gS}.
The values of $R$ and $\gamma_S$ are between the values obtained for the rapidity-integrated~($4\pi$) multiplicities measured in central Pb+Pb and in p+p collisions at the SPS energies~\cite{Vovchenko:2015idt,Begun:2018qkw}. The mid-rapidity values of $R$ and $\gamma_S$ resemble the Pb+Pb collisions, while forward rapidities resemble p+p. This is due to the fact that the largest amount of particles is produced at mid-rapidity, and their $dN/dy$ drop fast with increasing $y$, see Fig.~\ref{fig-dNdy}.
However, the radius parameter~(i.e. the system volume) for the most forward considered rapidity bin is still much larger than that in p+p collisions, see~\cite{Vovchenko:2015idt,Begun:2018qkw}.

The rapidity-integrated~($4\pi$) multiplicities give the system radius for the full hadron set (with $\Xi$ and $\Omega$) as $R=9.73\pm 0.46$~fm, while for the set without $\Xi$ and $\Omega$ the radius is $R=10.54\pm 0.48$~fm.
This corresponds to system volumes of $V=3864\pm543$~fm$^3$~(with $\Xi$ and $\Omega$), and $V=4898\pm672$~fm$^3$ (no $\Xi$ and $\Omega$).
The sum of the volumes obtained from the fits for the rapidity sub-bins is $V=4225\pm210$~fm$^3$~(full hadron set), and $V=5441\pm276$~fm$^3$~(no $\Xi$ and $\Omega$). Thus, the volume obtained for the fit of the rapidity integrated yields agrees with the volume obtained after the sum of the sub-bin volumes. It indicates that the performed $\Delta y$ binning is correct.
\\

The comparison of the rapidity and the energy scan within the UrQMD shows that the rapidity scan within the interval $0<y<3$ will deliver the systems, which have temperatures, baryo-chemical potential, volume, and strangeness content similar to the systems obtained at mid-rapidity.
The further increase of rapidity to $3<y<4.5$ will deliver the systems, which are at similar temperature, with larger baryon content, but with much less amount of strange particles, and in a smaller volume.
The rapid increase of $\chi^2/$DoF at rapidities $y\gtrsim 3.5$ suggests that the corresponding $y$ bins have less grounds to be treated in terms of temperature and baryo-chemical potential, as compared to the rapidity bins at $y\lesssim 3.5$.

\section{Summary}\label{Sec-Concl}
We conclude that the $dN/dy$ scan with Pb+Pb collisions at the $\sqrt{s_{NN}}=72$~GeV can be used to study systems which have similar temperatures, but at least a factor 2.5 larger baryon chemical potential in forward rapidity compared to mid-rapidity.
The rapidity scan at the AFTER@LHC covers the majority of the $\mu_B$ range accessible in the RHIC Beam Energy Scan program.\footnote{In the late stages of preparation of this paper an article~\cite{Karpenko:2018xam} appeared on arXiv. That paper covers a similar topic and reaches similar conclusions, but with a different method. Relativistic hydrodynamics is employed in \cite{Karpenko:2018xam} instead of a transport simulation. In contrast to \cite{Karpenko:2018xam}, here we additionally report on the degree of strangeness non-equilibrium and on the size of the system at freeze-out.}
Therefore, such a $dN/dy$ study in the AFTER@LHC project will provide a complementary approach to the QCD phase diagram studies, with all the benefits of a high-luminosity fixed-target experiment at the LHC.

The integrated luminosity for Pb+Pb collisions at AFTER@LHC is $L_{int} \sim 1.6 \, nb^{-1}$~\cite{Massacrier:2017oeq} per year, which corresponds to $\approx$11.5 billion minimum bias events. Given the particle yields predicted by the UrQMD, and assuming
even a very conservative value of the reconstruction efficiency of 1\%~\cite{Aamodt:2011zza, ABELEV:2013zaa} in the ALICE detector, the measurement precision  of all particle species (except for the heaviest and the infrequent $\bar{\Omega}$) yields over the whole rapidity range considered here will be limited only by the systematic uncertainties. Only the study of the $\bar{\Omega}$ baryon at the most forward rapidity bin could be a challenge, since the yield from the UrQMD model is low, and the success will depend on the $\Omega (\bar{\Omega})$ reconstruction efficiency available in the fixed-target mode. With a 4\% reconstruction efficiency, and assuming the signal-to-background ratio of 3:1~\cite{ABELEV:2013zaa} and the yield per event predicted by the UrQMD model, it will be possible to measure $\bar{\Omega}$ with 10\% relative uncertainty.
Thus, the studies proposed here are definitely within the reach of a single-year Pb+Pb program at the AFTER@LHC.
\\

\appendix
\setcounter{secnumdepth}{0}
\section{Appendix}

\begin{figure}[h!]
\includegraphics[width=0.49\textwidth]{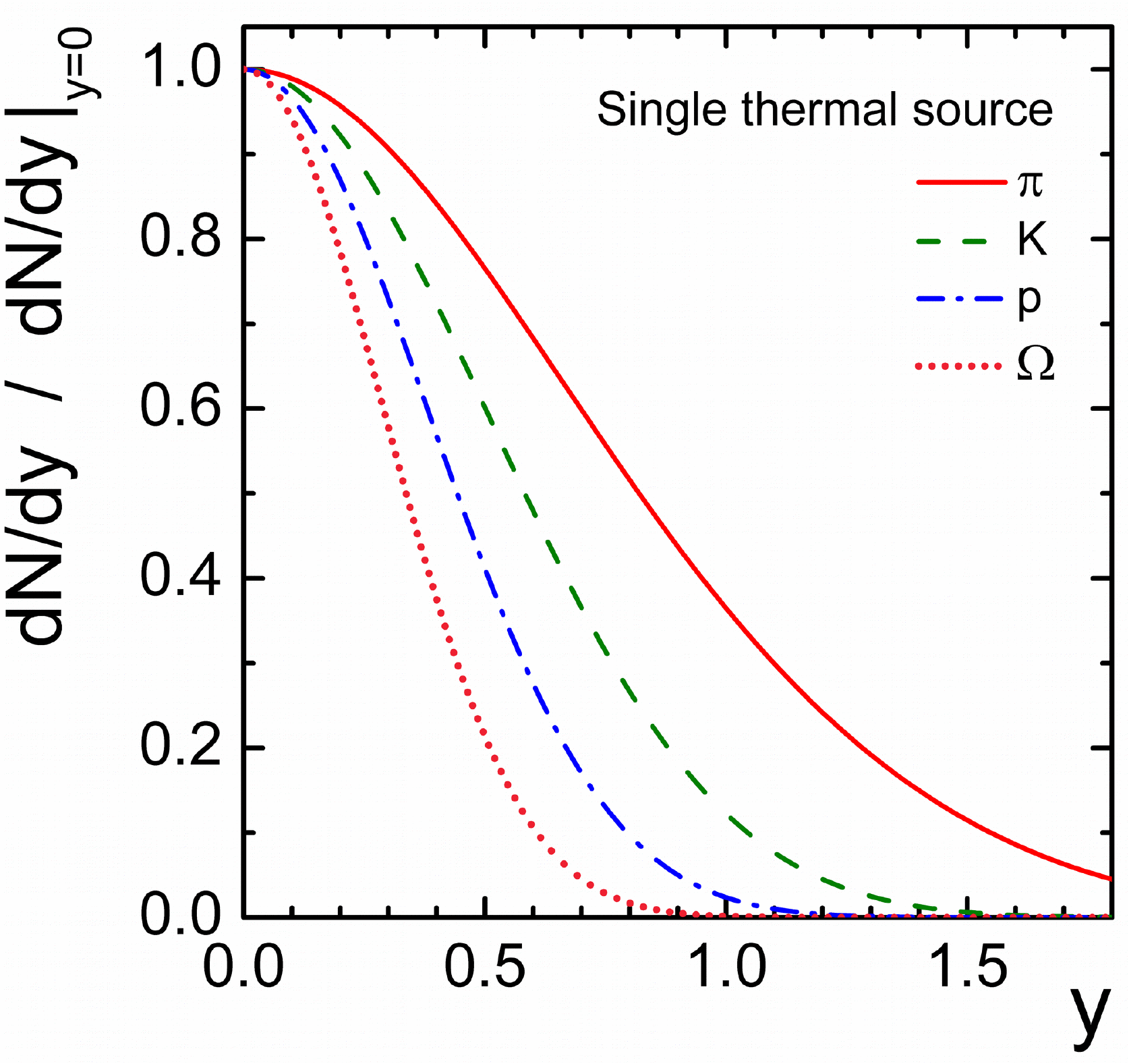}
\caption{Normalized rapidity distribution produced by a single thermal source at $T=140$~MeV.
}
\label{fig-TSource}
\end{figure}

Let us consider a single thermal source at rest~\cite{Schnedermann:1993ws}. The invariant momentum spectrum of particles radiated by this source is
 \eq{\label{eq:static}
 E\,\frac{d^3 N}{d^3p}
  ~=~ \frac{dN}{dy\,m_T\,dm_T\, d\phi}
   ~=~ \frac{gV}{(2\pi)^3}\,E\,e^{-(E-\mu)/T}~,
 }
where $E=\sqrt{m^2+p^2}=m_T\cosh(y)$ is the energy of a particle with mass $m$ and momentum $p$; $y$ and $\phi$ - rapidity and azimuthal angle of particl's 4-momentum; $m_T=\sqrt{m^2+p_T^2}$ and $p_T$ are, respectively, transverse mass and transverse momentum; $g$ and $\mu$ - the degeneracy factor of a particle and the chemical potential; $V$, $T$ - volume and temperature of the system. The integration over $\phi$ and $m_T$ in Eq.~\eqref{eq:static} yields the rapidity distribution
 \eq{\label{T-s}
 \frac{dN}{dy}
 &~=~ \int\frac{gV}{(2\pi)^3}\,E\,e^{-(E-\mu)/T}  m_T\,dm_T\, d\phi
  ~=~ \frac{V}{(2\pi)^2}\,\cosh(y)\,e^{\mu/T}\int_m^{\infty} e^{-m_T\cosh(y)/T} m_T^2\,dm_T
 \nonumber \\
 &~=~ \frac{gV}{(2\pi)^2}\,T^3\,e^{\mu/T} \left(
 \frac{2}{[\,\cosh(y)\,]^2} ~+~ \frac{m}{T}\,\frac{2}{\cosh(y)} ~+~ \frac{m^2}{T^2}
\right)e^{-\frac{m}{T}\cosh(y)}~.
 }
Equation~(\ref{T-s}) contains the dependence on particle's mass. We plot it in Fig.~\ref{fig-TSource} for $T=140$~MeV, $\mu=0$, and $m=140,~494,~938,~1672$~MeV, which corresponds, respectively, to pions, kaons, protons, and $\Omega$ baryons.
One may see that a single thermal source produces very different rapidity distributions for particles with different masses. The larger the mass - the narrower is the distribution.
Therefore, a selection of a narrow rapidity bin, and a fit of the particle multiplicities from this bin would produce a temperature and other parameters, which are different from those of the whole fireball.
A mid-rapidity cut would increase the  amount of heavy particles relative to light ones, while forward rapidity cut would decrease it. This behavior is qualitatively similar to the decrease of $\gamma_S$ with rapidity in our analysis of the UrQMD generated output, because strange particles are heavier than the non-strange.

However, the $dN/dy$ distributions that are produced in experiment~(and also in the UrQMD model) are not from a single thermal source, compare Fig.~\ref{fig-dNdy} and Fig.~\ref{fig-TSource}.
The rapidity distributions from a static thermal source have a Gaussian-like shape, with the full width at the half maximum $\Delta y_{_{\rm FB}}\simeq 1.6,~1.2,~0.9,~0.7$ for pion, kaon, proton and $\Omega$, correspondingly. In contrast, the UrQMD generated $dN/dy$ distributions have a wide plateau at mid-rapidity, $dN/dy\sim \textrm{const}$ on the interval $\Delta y \gg \Delta y_{_{\rm FB}}$.
This is the manifestation of a boost invariance of multiplicity distributions at midrapidity in high energy reactions. It also means that the wide $dN/dy$ distributions produced by an experiment or by the UrQMD can be approximated with a sum of thermal fireballs populating the rapidity axis, as was done for $\sqrt{s_{NN}}=200$~GeV at RHIC in~\cite{Biedron:2006vf,Broniowski:2007mu,Becattini:2007ci,Becattini:2007qr}.

Typically, the HRG model has been applied
in two extreme cases as far as the rapidity scan is concerned: in the full rapidity interval, or for a very narrow mid-rapidity.
We propose to treat each sub-bin as a separate fireball, and decide about the quality of this approximation by looking at the $\chi^2/$DoF values. Application of this procedure is much simpler than the fits of the $dN/dy$ with the sum of the fireballs. The change of the output HRG parameters in our case will indicate which particles are present at which rapidity, and to what extent the selected rapidity bin can be treated as a single fireball.

\acknowledgments

The authors thank M.~Gazdzicki, M.~I.~Gorenstein, K.~Grebieszkow, and J.~Steinheimer, for fruitful discussions. DK gratefully acknowledges inspiring discussions with Jean-Philippe Lansberg.
%

\bibliographystyle{h-physrev}
\bibliography{dNdyScan10}

\end{document}